\documentclass[global]{svjour}
\usepackage{graphicx}
\usepackage{times}
\journalname{Appl. Phys. B -- special issue `Nano optics' --}
\begin{document}

\title{Electromagnetic field correlations 
near a surface with a nonlocal optical response}

\author{Carsten Henkel$^1$%
\thanks{email: \tt Carsten.Henkel@physik.uni-potsdam.de}
\and
Karl Joulain$^2$}

\institute{$^1$Institut f\"ur Physik,
Universit\"at Potsdam, Germany,
\\
$^2$Laboratoire d'Etudes Thermiques,
Ecole Nationale Sup\'erieure
de M\'ecanique A\'eronautique, Poitiers, France
}

\date{23 March 2006}

\maketitle

\begin{abstract}
The coherence length of the thermal electromagnetic field
near a planar surface has a minimum value related to the
nonlocal dielectric response of the material. We perform
two model calculations of the electric energy density
and the field's degree of spatial coherence. Above a polar
crystal, the lattice constant gives the minimum coherence
length. It also gives the upper limit to the near field
energy density, cutting off its $1/z^3$ divergence.
Near an electron plasma described by the semiclassical
Lindhard dielectric function, the corresponding length scale
is fixed by plasma screening to the Thomas-Fermi length.
The electron mean free path, however, sets a larger scale
where significant deviations from the local description
are visible.
\end{abstract}

\titlerunning{The near field of a nonlocal surface}
\authorrunning{C. Henkel and K. Joulain}

\section{Introduction}

Thermal electromagnetic radiation in vacuum, as described by the celebrated
blackdody spectrum discovered by Max Planck \cite{Planck00}, is
usually taken as a typical example of incoherent radiation. This is
not quite true, however: if the radiation is detected at a given
frequency, it is spatially coherent on a scale set by the
wavelength \cite{Rytov3,Gori94}. When one approaches
a macroscopic object, the radiation spectrum and its coherence is 
profoundly changed, depending
on the properties of the object. For example, if the object supports
resonant modes like surface plasmon polaritons, the field is coherent
across the propagation length of these modes 
\cite{Greffet99}. The opposite case is possible as well:
the coherence length becomes comparable to the observation distance,
much smaller than the wavelength, close to an absorbing object with a 
local dielectric function \cite{Henkel00b}. It has been suggested
already by Rytov and colleagues that this behaviour is an artefact
because at some small scale, nonlocal effects must come into play
\cite{Rytov3}. This is what we discuss in this paper in a 
semi-quantitative way. We use two models for nonlocal dielectric
functions and identify the scale for the field's coherence length
using explicit asymptotic formulas. A nonlocal dielectric response
is of primary importance for semiconductor quantum wells, see for
example Ref.\cite{Savasta99}, but the issue of spatial coherence 
has not been analyzed in this context, to our knowledge.  

We focus on the spatial coherence of the electromagnetic
field at nanometer distance in the vacuum (medium 1) above a 
solid material (medium 2).  
We chose a planar geometry which  
is sufficiently simple to allow for an essentially analytical
description, thus avoiding the need for extensive numerics.  On the
other hand, many effects have been discussed in this setting:
the fluorescence rate of molecules near metals and thin films
\cite{Chance78}, 
scanning near-field microscopy of sub-wavelength objects deposited 
on a substrate \cite{Dunn99},
the momentum exchange between a tip and a sample (related to the Casimir 
force, see, e.g., \cite{Mohideen02})
and the energy exchange between a tip and a sample
\cite{Xu94,Pendry99b,Mulet01,Kittel05}.

\section{Basic notation}

\subsection{Field correlations}

The spatial coherence of the electric field is determined by the 
two-point expectation value \cite{MandelWolf}
\begin{equation}
\langle E_{i}( {\bf r}_{1}, t_{1} ) 
E_{j}( {\bf r}_{2}, t_{2} ) \rangle
= \int\!\frac{ {\rm d}\omega }{ 2 \piÊ}
{\cal E}_{ij}( {\bf r}_{1}, {\bf r}_{2}; \omega )
{\rm e}^{ {\rm i} \omega ( t_{1} - t_{2} ) }
,
    \label{eq:coh-spectrum}
\end{equation}
where the average is taken in a stationary statistical ensemble
(thermal equilibrium in the simplest case).  We focus in the following
on the cross-correlation spectrum ${\cal E}_{ij}( {\bf r}_{1}, 
{\bf r}_{2}; \omega )$ and a frequency in the infrared to visible 
range. Far from any sources and in global equilibrium, the 
corresponding wavelength $\lambda = 2\pi c / \omega$ sets the scale 
for the field's spatial coherence length: the cross-correlations tend 
to zero if the distance $|{\bf r}_{1} - {\bf r}_{2}|$ exceeds $\lambda$.
In the vicinity of a source, the coherence length $\ell_{\rm coh}$ 
significantly differs from $\lambda$, as Henkel and co-workers 
have shown previously
\cite{Henkel00b}, and it changes with the observation point.

The spectrally resolved electric energy density is given by the trace
\begin{equation}
    u_{E}( {\bf r}; \omega) 
    = \frac{\varepsilon_{0}}{2} \sum_{i}
    {\cal E}_{ii}( {\bf r}, {\bf r}; \omega ),
    \label{eq:def-el-energy}
\end{equation}
and its value in thermal equilibrium allows to define an electric, 
local density of states, as discussed in more detail by Joulain and 
co-workers \cite{Joulain03}. The normalized tensor
\begin{equation}
    c_{ij}( {\bf r}_{1}, {\bf r}_{2}; \omega ) = 
    \frac{ \frac12 \varepsilon_{0} 
    {\cal E}_{ij}( {\bf r}_{1}, {\bf r}_{2}; \omega ) 
    }{ \sqrt{ u_{E}( {\bf r}_{1}; \omega) 
    u_{E}( {\bf r}_{2}; \omega)} }
    ,
    \label{eq:degree-of-coherence}
\end{equation}
to be considered below, allows to introduce a spatial degree of coherence.
In the following, we call a ``coherence function'' both,
${\cal E}_{ij}( {\bf r}_{1}, {\bf r}_{2}; \omega )$
and Eq.(\ref{eq:degree-of-coherence}). Definitions for a degree of 
polarization based on this $3\times 3$ matrix (with ${\bf r}_{1} = 
{\bf r}_{2}$) have been put forward as well, see \cite{Friberg02,Wolf04a}.
For the sake of simplicity, we suppress the frequency arguments
in the following.

\subsection{Planar surface with local response}

In a previous paper, Henkel and co-workers have shown that in the 
vacuum 
above a planar dielectric surface at temperature $T$, described by 
a local permittivity $\varepsilon_{2}$,
the spatial coherence function  is of the form 
\cite{Henkel00b} (see also \cite{Girard00a,HenkelHabil})
\begin{equation}
    {\cal E}_{ij}( {\bf r}_{1}, {\bf r}_{2} ) = 
    \frac{ \Theta( \omega, T ) }{ 2\pi\varepsilon_{0} \omega \, \tilde r^{5} }
    {\rm Im}\frac{ \varepsilon_{2} - 1 }{ 
    \varepsilon_{2} + 1 }
    \left( 
    \begin{array}{ccc}
	\tilde r^2 - 3 \rho^2 & 0 & 3 \rho ( z_{1} + z_{2} ) 
	\\
	0 & \tilde r^2 & 0
	\\
	-3 \rho( z_{1} + z_{2} ) & 0 & 3 ( z_{1} + z_{2} )^2 - \tilde r^2 
    \end{array}
    \right)
    \label{eq:local-xnf}
\end{equation}
where $\Theta(\omega,T)=\hbar\omega/(e^{\hbar\omega/kT} - 1)$. 
We assume that the field is observed in vacuum (relative
permittivity $\varepsilon_{1} = 1$).
The surface is given by $z = 0$.
We have chosen the $x$-axis such that ${\bf r}_1 - {\bf r}_2$ lies in 
the $xz$-plane and $\rho = x_{1} - x_{2}$. The quantity 
$\tilde r^2 = \rho^2 + (z_{1} + z_{2})^2$ is the (squared) 
distance between ${\bf r}_{1}$ and the image point of 
${\bf r}_{2}$ across the interface.

Eq.(\ref{eq:local-xnf}) applies to leading order
when both distances $z_{1}$, $z_{2}$ are much 
smaller than the wavelength $\lambda$; for other regimes and higher 
order corrections, see Ref.\cite{Henkel00b} and, at $\rho = 0$, 
Ref.\cite{Scheel99a}.
In the following, we focus on the correlation function
at a constant height $z = z_1 = z_2$ and discuss its dependence on 
the lateral separation $\rho$; note that $\rho$ can be positive or negative.
The normalized coherence function~(\ref{eq:local-xnf}) is qualitatively 
similar to a 
Lorentzian: the $yy$-component, for example, follows 
a law $\sim [4z^2 + \rho^2]^{-3/2}$. The spatial coherence length is 
thus equal to $z$, and decreases without apparent limitation as the surface 
is approached. The electric energy density derived 
from~(\ref{eq:local-xnf}) diverges like $1/z^3$:
\begin{equation}
u_{E}( z ) = [ \Theta( \omega, T) / ( 8\pi \, z^3)]
{\rm Im}[( \varepsilon_{2} - 1 )/(
\varepsilon_{2} + 1 )]
.
\label{eq:ue-local}
\end{equation}
Both points have been 
noted by Rytov and co-workers \cite{Rytov3}, who have also argued that 
this unphysical result is due to the assumption of a local dielectric 
response down to the smallest scales. A cutoff would occur naturally 
in a non-local treatment or taking into account the atomistic
structure of the material. This is what we show here in detail, using 
two different model calculations. Doing this, we also provide a 
basis for the phenomenological cutoff introduced 
recently by Kittel and co-workers \cite{Kittel05} in the context of 
heat transfer from a hot, sharp tip into a cold, planar substrate.

\subsection{Overview}

We will use two models to calculate the coherence function.
In both, we focus, as mentioned before, on the fields near a planar 
surface and compute the field correlations in the vacuum above 
it, at sub-wavelength distances.

The first model is based on the fluctuation electrodynamics 
introduced by Rytov and co-workers \cite{Rytov3} where the sources of 
the field are described by fluctuating polarization currents below the 
surface. This approach relies on the 
fluctuation-dissipation theorem that links the spectrum of the
polarization current to the dielectric function of the material below 
the surface. 
We argue that the currents are spatially correlated on a 
scale equal to or larger than the material's microscopic lattice constant.
We then show that the radiation generated outside the surface shows 
a minimum coherence length given by this scale; this cuts off the 
divergences appearing in a local description of the material, as 
noted in Refs.\cite{Rytov3,Kittel05}.
This model can be applied to polar ionic crystals in the frequency 
domain where the dielectric response is dominated by phonon-polariton 
resonances.
It can also cover a non-equilibrium situation 
where the surface is heated to a different temperature or shows weak 
temperature gradients \cite{Polder71,Henry96}. 

The second model describes the dielectric response of an electron 
plasma and applies to the plasmon-polariton resonances occurring in 
metals. We use here directly the fluctuation-dissipation theorem for 
the electric field \cite{Callen51,Eckhardt82}, restricting ourselves 
to a field-matter system 
in ``global equilibrium''. The coherence function is determined 
by reflection coefficients from the surface for which we take the 
Lindhard form, taking into account the non-local response of the 
electron plasma. It is shown that the field's coherence length is limited 
by the Thomas-Fermi screening length, but significant deviations
from the local description occur already on the (typically larger) 
scale of the electron mean free path.

\section{Polar crystal}

\subsection{Current correlations}

We assume here that the fluctuating currents that generate the 
radiation field, are correlated
below a certain distance $l$.  Above this distance, the medium response 
can be considered as local.
A lower limit for $l$ is certainly the lattice period $a$: at scales
smaller than $a$, the concept of a continuous material characterized
by a dielectric constant does not make sense any more.

In this situation, the cross correlation spectrum of the
fluctuating currents, as given by the fluctuation-dissipation theorem, 
is no longer delta-correlated in space.  We choose here to smoothen the 
spatial delta fonction into a gaussian.  The fluctuation-dissipation theorem
for the currents thus takes the form
\begin{equation}
\label{FDTnonloc}
\left<j_k^*({\bf r}_1,\omega)j_l({\bf r}_2,\omega')\right> = 
2\omega\varepsilon_0
{\rm Im}[ \varepsilon( \bar \mathbf{r} ) ]
%{\textstyle\frac12}(\mathbf{r}_1 + \mathbf{r}_2))] 
\frac{ e^{-({\bf r}_1 - {\bf r}_2)^2/l^2} }{ \pi^{3/2} l^3 }
\Theta(\omega,T)
\delta_{kl}
\delta(\omega - \omega')
,
\end{equation}
where $\bar\mathbf{r} = \frac12(\mathbf{r}_1 + \mathbf{r}_2)$. The 
gaussian form for the spatial smoothing is chosen for convenience; 
another 
functional dependence, e.g.\ the model put forward by Kliewer and 
Fuchs \cite{KliewerFuchs}, will lead to qualitatively similar results.

\subsection{Transmitted field}

We then write the cross correlation spectrum for the electric field 
in terms of Green functions and the currents.  We use the convention 
\begin{equation}
E_i({\bf r},\omega) = 
{\rm i}\mu_0\omega
\int d^3{\bf r'} \,
\sum_{k}
G_{ik}({\bf r},{\bf r}';\omega) 
j_k({\bf r}', \omega)
.
\end{equation}
To proceed further in the calculation, the Green function is written as a
Weyl plane wave expansion (\cite{Sipe84} and appendix).  In the
present case, the Green function relates the current on one side of an
interface to the electric field on the other side of the interface.
It depends on the Fresnel transmission coefficients through this
interface. 

Using (\ref{FDTnonloc}) and integrating over the half-space filled with 
the dielectric, one obtains
\begin{eqnarray}
\label{integK}
{\cal E}_{ij}( {\bf r}_{1}, {\bf r}_{2} ) = 
2\mu_0\Theta(\omega,T) \omega
\int_0^{2\pi}\frac{ {\rm d}\theta }{ 2\pi }
\int_0^\infty \frac{K \,{\rm Re}(\gamma_2) {\rm d}K}{2 \pi \, |\gamma_2|^2}\\
\times e^{-i K \rho \cos\theta}
e^{-2 {\rm Im}(\gamma_1)z}e^{-K^2l^2/4}
e^{-{\rm Re}(\gamma_2)^2l^2/4}
g_{ik}^*({\bf K})g_{jk}({\bf K})
\nonumber
\end{eqnarray}
In the preceding equation, the wavenumber in the medium $i = 1, 2$ is 
${\bf k}_i = ({\bf K},\gamma_i)$ where
${\bf K}=K \cos\theta \,{\bf e}_x + K \sin\theta \,{\bf e}_y$ 
and $\gamma_i^2=\epsilon_i (\omega/c)^2-k^2$. The tensor 
$g_{ij}({\bf K})$ is related to the Green tensor and defined in the 
Appendix.  

The cross-spectral correlation function depends on four
characteristic lengths: the wavelength $\lambda$, the distance to the
interface $z$, the locality distance $l$ and the separation $\rho$
between the field points. The latter is the variable
considered in our problem.  At the wavelengths we work with, we always
have $l\ll \lambda$.  When $z$ is larger than $\lambda$ (in the far field), 
the factor $e^{-2 {\rm Im}(\gamma_1)z}$ actually limits the
integration over $K$ to $0 \le K \le \omega/c$, i.e., to propagating
waves.  The cross-spectral correlation function, in this regime,
drops to $0$ when $\rho$ exceeds $\lambda/2$, as in the blackbody 
radiation field. In the intermediate regime $l\ll z\ll\lambda$,
the integral is dominated by the range $\omega/c \ll K \ll 1/l$,
where the exponentials containing $l$ are close to unity. 
Hence, the results of 
Ref.\cite{Henkel00b} are recovered.  
Finally, when $z\ll l$, $e^{-2{\rm Im}(\gamma_1)z}$ and
$e^{-{\rm Re}(\gamma_2)^2 l^2/4}$ both approach unity in the 
relevant range $|\sqrt{\varepsilon_{2}}|\omega/c \ll K < 1/l$.  This
is the regime we discuss in more detail in the following.

We note in passing that we use our calculation is based on the
solution to the transmission problem valid for a local
medium.  Actually, this solution applies when the wave vector $K$ is 
smaller than $1/l$ when the medium can be described as homogeneous.
But from~(\ref{integK}) one sees that whatever
the values of $z$, there is anyway a cut-off in the integration over
$K$ at
approximately $1/l$.  Therefore, one might consider that the local
expression of the Fresnel coefficients remains valid.  We believe that 
our model, even if it not rigorously accurate, is useful
in view of the insight one gains from the analytic result.

\subsection{Asymptotics and discussion}

\begin{figure}
\begin{center}
\includegraphics*[width=100mm]{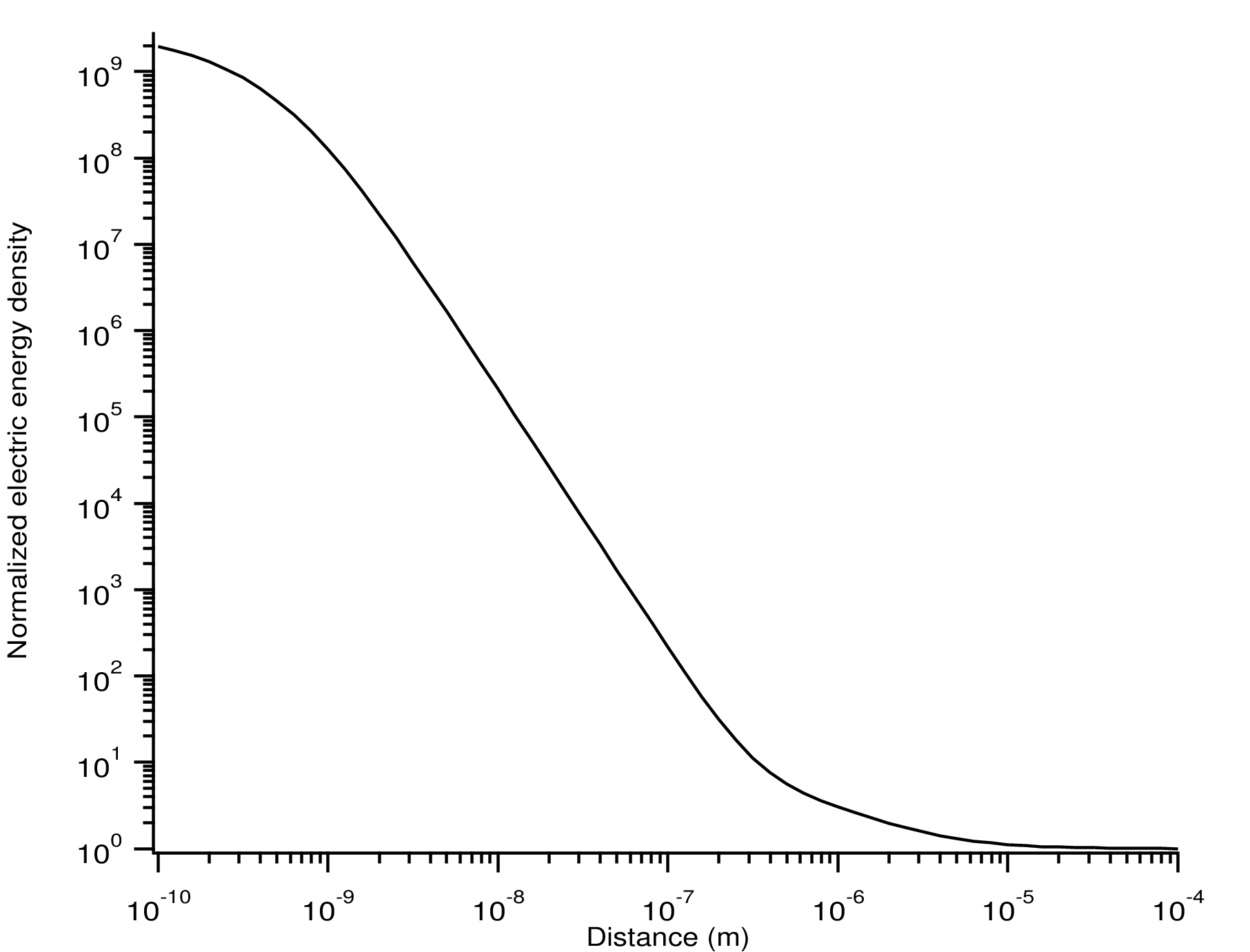}
\end{center}
\caption{Normalized electric energy density above a surface of silicon
carbide vs.\ the distance $z$ to the surface.  The electric energy
density is normalized to the electric energy density in the far field.
The locality scale is taken as $l = 1$nm. The SiC permittivity is 
described by an oscillator model in the visible-infrared part of 
the spectrum \cite{Palik}.}
\label{fig:saturation}
\end{figure}

Using the limit of $g_{ij}( {\rm K} )$ for large $K$, we obtain
from~(\ref{integK}) the following asymptotic expression for the cross
spectral correlation tensor
\begin{eqnarray}
\label{eq:Eij-polar}
&&
{\cal E}_{ij}( {\bf r}_{1}, {\bf r}_{2} )
\approx
\frac{8\Theta(\omega,T) {\rm Im}(\varepsilon_{2})}
{\varepsilon_0\pi\omega|\varepsilon_{2} + 1|^2 l^3}\\
&&
\times
\left(
\begin{array}{ccc}
    \frac{\sqrt{\pi}}{2}[M_{3/2} - \frac{3}{4}\frac{\rho^2}{l^2}M_{5/2}]    
    & 0 
    & -2\frac{\rho}{l}e^{-\rho^2/l^2}    
    \\
    0 
    & \frac{\sqrt{\pi}}{2}[M_{3/2}+\frac{3}{4}\frac{\rho^2}{l^2}M_{5/2}] 
    & 0  
    \\
    2\frac{\rho}{l}e^{-\rho^2/l^2}  
    & 0 
    & \sqrt{\pi}M_{3/2}
\end{array}
\right)
,
\nonumber
\end{eqnarray}
where $M_{3/2} = M(\frac{3}{2},1,-\frac{\rho^2}{l^2})$ and $M_{5/2} =
M(\frac{5}{2},3,-\frac{\rho^2}{l^2})$, and $M(a,b,z)$ is the confluent
hypergeometric function \cite{Abramowitz}.  
When $\rho \ll l$, $M_{3/2}$ and $M_{5/2}$ both approach unity.
Putting $\rho=0$ in the cross-spectral
correlation tensor and taking the trace, we get the electric energy
density versus $z$:
\begin{equation}
z \ll l: \qquad
    u_{E}( z ) = \frac{2 \Theta( \omega, T )}{\pi^{1/2}
    \omega \, l^3}
    {\rm Im} \frac{\varepsilon_{2} - 1}
    {\varepsilon_{2} + 1}.
    \label{eq:energy-density-xnf}
\end{equation}
It appears (see Fig.\ref{fig:saturation}) that it saturates at short $z$ to a
quantity that only depends on $l$ as $1/l^3$: the non-locality scale
$l$ thus sets the ultimate length below which the field properties 
are ``frozen'' to their value for $z \approx l$.

\begin{figure}
\begin{center}
\includegraphics*[width=80mm]{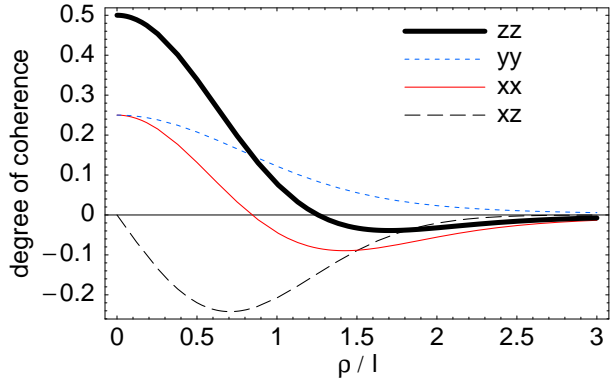}
\end{center}
\caption{Normalized spatial coherence function vs.\ lateral 
separation $\rho$ in units of the nonlocality scale $l$.
The nonzero components in Eq.(\ref{eq:Eij-polar}) are plotted 
and normalized to the trace of the 
coherence tensor.}
\label{fig:cij-polar}
\end{figure}

When $\rho \gg l$, all the components of the correlation
tensors drop to zero, see Fig.\ref{fig:cij-polar}.
This decrease is exponentially fast for the
$xz$ and $zx$ components.  For the other components, the
asymptotic behaviour for large $\rho$ simply scales like $1/\rho^3$ 
and does not depend on $l$ anymore.  This follows from the large 
argument asymptotics 
$M_{3/2}\approx\frac{-1}{2\sqrt{\pi}} \frac{l^3}{\rho^3}$ and
$M_{5/2}\approx\frac{2}{\sqrt{\pi}} \frac{l^5}{\rho^5}$.  Note that in
this case, we recover an algebraic decay similar to the local medium 
case given in Eq.(\ref{eq:local-xnf}).

To summarize this section, we have shown that when we take into
account the non-local nature of matter by introducing a locality
length $l$ for the sources of the field,
the correlation length is about $l$ when the distance to
the interface $z<l$. In this regime, the energy density saturates to a
value given by the electrostatic energy density expression taken in
$z=l$.

\section{Nonlocal plasma}

We consider in this section another simple situation where the field 
correlation function can be calculated fairly easily. Restricting 
ourselves to a field in thermal equilibrium between field and surface,
we use directly the fluctuation-dissipation theorem for the field.
The relevant information is thus encoded in the electric Green tensor 
(i.e., the field's response function). The Green tensor contains a 
part due to the reflection from the surface that is actually 
dominating in the sub-wavelength distance regime we are interested in. 
We first review the corresponding reflection coefficients for an 
electron plasma, taking into account the finite response time of the 
electrons and their scattering. These two effects
make the plasma behave like a nonlocal medium and give rise to the
so-called anomalous skin effect.
We then discuss the large-wavevector 
asymptotics of the reflection coefficients and the corresponding 
limits on the spatial coherence function. It turns out that the 
scattering mean free path is one key quantity that limits the 
coherence length at short observation distances.

\subsection{Reflection coefficients}

We focus here on the electronic contribution to the dielectric 
function and describe the background ions, interband absorption etc.\,
by a local permittivity 
$\varepsilon_{b}$ to avoid unnecessary complications.
As is well known \cite{AshcroftMermin}, the dielectric function of a 
bulk plasma is actually 
a tensor with two distinct spatial Fourier coefficients, a 
``longitudinal'' $\varepsilon_{l}( {q} )$ and 
a ``transverse'' $\varepsilon_{t}( {q} )$ where $q$ is the modulus
of the wavevector. (As before, we suppress 
the frequency dependence for simplicity.) 
The fields outside the metal surface are characterized by the 
reflection coefficients $r_{s,p}( {K} )$ that depend only on
the magnitude $K = |{\bf K}|$ of 
the incident wavevector projected onto the interface.
Out of the two polarizations s and 
p, we need in the following only $r_{p}( {K} )$ in the (non-retarded)
regime $K \gg \omega / c$. 
This coefficient is given, e.g., in the review paper 
by Ford and Weber \cite{Ford84}: 
\begin{equation}
    r_{p}( K ) = \frac{ 1/Z_{p}(K) - 1 }{ 1/ Z_{p}(K) 
    + 1}
    \label{eq:rp-impedance}
\end{equation}
We use here a dimensionless surface impedance $Z_{p}(K)$ that reads
in the non-retarded limit 
\begin{equation}
    Z_{p}( K ) = 4 K \int\limits_{0}^{\infty}\frac{ {\rm d}k_z }{ 
    2 \pi } \frac{ 1 }{ q^2 \varepsilon_{l}( {q} ) },
    \qquad q^2 = K^2 + k_z^2
 ,
    \label{eq:Zp-eps-l}
\end{equation}
it involves the longitudinal dielectric function only for which we take
the Lindhard formula \cite{Ford84,Kliewer68}
\begin{eqnarray}
    \varepsilon_{l}( {q} ) &=& \varepsilon_{b} + 
    \frac{ 3 \Omega^2 }{ \omega + {\rm i} \nu }
    \frac{ u^2 f_{l}( u ) }{ \omega + {\rm i} \nu f_{l}( u ) } 
    \label{eq:eps-l-Lindhard}
    \\
    u &=& \frac{ \omega + {\rm i}\nu }{ q v_{F} } \equiv \frac{ 1 }{ 
    q \ell }
    \label{eq:def-complex-ell}
    \\
    f_{l}( u ) & = & 1 - \frac{ u }{ 2 } \log\left( \frac{u + 1}{u - 
    1} \right)
    .
    \label{eq:fl-Lindhard}
\end{eqnarray}
The plasma frequency is given by $\Omega^2 = n e^2 / 
(m \varepsilon_{0})$ with $n, -e, m$ the electron density, charge, and
mass, respectively.

From the nonlocal 
permittivity~(\ref{eq:eps-l-Lindhard}--\ref{eq:fl-Lindhard}),
two characteristic length scales can be read off: the mean free path
$l_{\rm mfp} = v_{F} / \nu$ and $v_{F} / \omega$, the maximum distance 
over which an electron at the Fermi energy can move ballistically during
one period of the applied electric field.  In the following, we use
the complex length $\ell = v_{F} / (\omega + {\rm i} \nu)$ defined
in~(\ref{eq:def-complex-ell}) to simplify the notation.

The Lindhard formula, 
Eqs.~(\ref{eq:eps-l-Lindhard}--\ref{eq:fl-Lindhard}),
is based on a
semiclassical description of the electron gas (classical particles
with Fermi statistics) with a damping rate $\nu$ and a velocity 
$v_{F}$ at the Fermi energy. 
This description is valid as long as $q$
is much smaller than the Fermi wave vector $k_{F} = m v_{F} / \hbar$.
Our model thus applies reasonably well to a ``clean metal'' where the
mean free path is much longer than the Fermi wavelength, and to distances 
above $1/k_{F}$ (typically a few \AA). Ref.\cite{Ford84} gives 
a more general dielectric function that covers the regime $q \ge 
k_{F}$ as well.

\subsection{Coherence function}

The fluctuation-dissipation theorem for the electric field, combined 
with the Green tensor describing the reflection from a planar surface,
gives the following integral representation for the field's coherence 
function:
\begin{equation}
    {\cal E}_{ij}( {\bf r}_{1}, {\bf r}_{2} ) =
    \mu_{0} \omega \Theta( \omega, T ) 
    \int\limits_{0}^{\infty}\frac{ K {\rm d} K}{ 2\pi }
    \sum_{\mu = {\rm s, p}} {\cal C}_{ij}^{(\mu)}( K \rho ) 
    \, {\rm Re}\,\frac{ r_{\mu}( K ) \, {\rm e}^{ 2 {\rm i} \gamma_1 z }
    }{ \gamma_1 }
    \label{eq:coh-Im-r-0}
\end{equation}
with $\gamma_1 = (\omega^2/c^2 - K^2)^{1/2}$ (${\rm Im}\,\gamma_{1} 
\ge 0$).
For more details, 
see for example \cite{Agarwal75a,Dorofeyev02a}. We have omitted the 
free-space part of the Green tensor that gives the same result as for 
the blackbody field. This part actually becomes negligible compared to 
the surface part given here if we focus on the sub-wavelength regime,
$z_{1} = z_{2} = z \ll \lambda$: the integration domain 
$\omega/c \le K < \infty$ (which is absent in the free-space field)
then makes the dominant contribution to the integral.

The tensors ${\cal C}_{ij}^{(\mu)}( K \rho )$ in~(\ref{eq:coh-Im-r-0})
depend on the lateral (signed) distance $\rho = x_{1} - x_{2}$, as
introduced after Eq.(\ref{eq:local-xnf}).  In p-polarization, it is
given by
\begin{equation}
    {\cal C}^{({\rm p})}( K \rho ) = 
    \frac{ K^2 c^2 }{ 2 \omega^2 } \left(
    \begin{array}{ccc}
	J_{0} - J_{2} & 0 & 2 J_{1} 
	\\
	0 & J_{0} + J_{2} & 0
	\\
	-2 J_{1} & 0 & 2 J_{0} 
    \end{array}
    \right)
    ,
    \label{eq:C-p-tensor}
\end{equation}
involving the Bessel functions $J_{n} = J_{n}( K \rho )$, $n = 0, 1, 
2$. A similar 
expression applies in s-polarization. We can focus, for short 
vertical distances,
on the range $\omega/c \ll K$, expand 
the reflection coefficients and find that $|r_{s}| \ll 
|r_{p}|$; hence, the s-polarization is neglected in the following. 
This also justifies our taking the non-retarded
limit of the reflection coefficient~(\ref{eq:rp-impedance}).
To the same accuracy, we approximate $\gamma_1 \approx i |K|$.
Finally, the correlation tensor becomes
\begin{equation}
    {\cal E}_{ij}( {\bf r}_{1}, {\bf r}_{2} ) =
    \mu_{0} \omega \Theta( \omega, T ) 
    \int\limits_{0}^{\infty}\frac{ {\rm d} K}{ 2\pi }
    {\rm e}^{ - 2 K z }
    \sum_{\mu = {\rm s, p}} {\cal C}_{ij}^{(\mu)}( K \rho ) 
    \, {\rm Im}\,r_{\mu}( K ) 
    .
    \label{eq:coh-Im-r}
\end{equation}
We anticipate from the integral representation~(\ref{eq:coh-Im-r}) 
that the wave-vector dependence of ${\rm Im}\, r_{p}( K )$
determines the spatial coherence length: if $K_{c}$ is the scale on 
which ${\rm Im}\,r_{p}( K ) \to 0$, we expect that the 
divergence of the energy density is smoothed out for $z \ll 1/K_{c}$ 
and that the lateral coherence length remains finite:
$\ell_{\rm coh} \sim 1/K_{c}$ for $z \le 1/K_{c}$.

\subsection{Local medium}

Let us illustrate first how the Lindhard reflection coefficient 
reduces to its local form (the Fresnel formula). If the 
$q$-dependence of $\varepsilon_{l}( q ) $ can be neglected,
writing $\varepsilon_{l}( q ) \to 
\varepsilon_{\rm loc}$, 
the surface impedance~(\ref{eq:Zp-eps-l}) integrates to 
$Z_{p} \to 1 / \varepsilon_{\rm loc}$. Eq.(\ref{eq:rp-impedance}) then 
recovers the reflection coefficient for electrostatic images,
$r_{p} \to (\varepsilon_{\rm loc} - 1)/(\varepsilon_{\rm loc} + 
1)$ which is the large $K$ limit of the Fresnel formula for 
transverse magnetic (TM or p) polarization.
The integration of the Bessel functions and exponentials over $K$ in 
Eq.(\ref{eq:coh-Im-r}) can be 
carried out, and we get Eq.(\ref{eq:local-xnf}) with its unphysical 
$1/z^3$ divergence.

The same divergence would be obtained here from the background 
permittivity $\varepsilon_{b}$ that we assume local. To focus 
on the nonlocal contribution from the electron plasma, we consider the 
regime where $\varepsilon_{b}$ is real so that the leading-order,
local contribution analogous to Eq.(\ref{eq:local-xnf}) vanishes.
% , proportional 
% to
% ${\rm Im}[( \varepsilon_{b} - 1 )/(
% \varepsilon_{b} + 1 )]$, vanishes.
%[Next-to-leading order: see a Stefan Scheel paper.]

\subsection{Nonlocal reflection coefficient}

To get a qualitative insight into the impact of nonlocality,  
we perform an asymptotic analysis of the dielectric 
function~(\ref{eq:eps-l-Lindhard}--\ref{eq:fl-Lindhard}):
\begin{equation}
    \varepsilon_{l}( {q} ) \approx \left\{
    \begin{array}{ll}
	\displaystyle
    \varepsilon_{b} - \frac{ \Omega^2 }{ \omega( \omega + {\rm i} \nu) }
    \left[ 1 + \left( q \ell 
    \right)^2 \left( \frac35 + \frac{ {\rm i} \nu }{ 3 \omega } \right)
    \right]
    ,
    & \quad | q \ell | \ll 1
    \\[2ex]
    \displaystyle
    \varepsilon_{b} \left( 1 + \frac{ 1 }{ q^2 \Lambda^2 } \right)
    + \frac{ {\rm i} C }{ q^3 }
    ,
    & \quad | q \ell | \gg 1
    \end{array}
    \right.
    \label{eq:eps-l-asymp}
\end{equation}
where 
\begin{equation}
    \Lambda = \sqrt{ \varepsilon_{b} v_{F}^2 / (3\Omega^2)}
    \label{eq:def-Lambda}
\end{equation}
is the Thomas-Fermi length that provides another length scale,
and we use the notation $C = 3\pi \omega \Omega^2 / v_{F}^3$. 
We recall that $\ell$ is the complex characteristic length
defined in~(\ref{eq:def-complex-ell}). 
Note that for small $q$, we recover the usual, local Drude expression
for an electron plasma
\begin{equation}
    \varepsilon_{\rm loc} = \varepsilon_{b} - 
    \frac{ \Omega^2 }{ \omega ( \omega + {\rm i} \nu ) }
    .
    \label{eq:Drude-metal}
\end{equation}
At large $q$, one gets the dielectric function for Thomas-Fermi screening
\cite{AshcroftMermin}
with a screening length on the order of $v_{F} / \Omega$ plus an imaginary 
correction.

From the integral~(\ref{eq:Zp-eps-l}) for the surface 
impedance, we find that the typical wavenumber is of the 
order of $q \ge K$. Hence the two limits quoted above translate into 
the following asymptotics of the reflection coefficient, after 
performing the integrations,
\begin{equation}
    {\rm Im}\, r_{p}( K ) \approx \left\{
    \begin{array}{ll}
	\displaystyle
	{\rm Im} \frac{ \varepsilon_{loc} - 1 }{ 
	\varepsilon_{loc} + 1 }
	,
	& \quad | K \ell | \ll 1,
    \\[2ex]
    \displaystyle
    \frac{ 4 }{ 3 \varepsilon_{b}^2 }
    \frac{ C K \Lambda^4 g( K \Lambda ) }{ 
    \left| 1 + K / ( \varepsilon_{b} 
    \sqrt{ K^2 + 1/\Lambda^2 } ) 
    \right|^2 }
    ,
    & \quad | K \ell | \gg 1
    .
    \end{array}
    \right.
    \label{eq:r-p-asymp}
\end{equation}
The dimensionless function $g( K \Lambda )$ is the integral
\begin{equation}
g(K\Lambda) = \int\limits_0^\infty\!\frac{ {\rm d}t }{
\sqrt{ (K\Lambda)^2 + t^2 } [ (K\Lambda)^2 + 1 + t^2 ]^2 }
.
\label{eq:def-g}
\end{equation}
This can be evaluated in closed, but barely instructive form 
involving a hypergeometric function; its limiting behaviour is 
\begin{equation}
    \begin{array}{rcll}
	g( K \Lambda ) &\approx& \ln( 1/ K \Lambda ) + \ln 2 - \frac12 
	& \mbox{for } K \Lambda  \ll 1,
	\\
	g( K \Lambda ) & =&  \frac23 ( K \Lambda  )^{-4}
	& \mbox{for } K \Lambda \gg 1
	.
    \end{array}
    \label{eq:dimless-g}
\end{equation}
The first line applies to the intermediate case $1/|\ell| \ll K \ll
1/\Lambda$, the second one to the regime
$K \gg 1/\Lambda, 1/|\ell|$.
In both cases, Eq.(\ref{eq:r-p-asymp})
implies that $|{\rm Im}\, r_{p}( K )| \ll 1$. 

\begin{figure}
\includegraphics*[width=70mm]{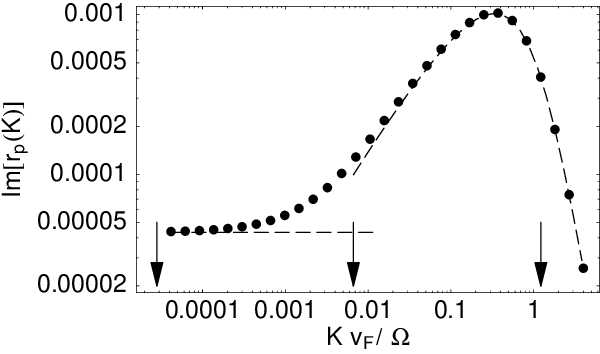}
\hspace*{5mm}
\includegraphics*[width=70mm]{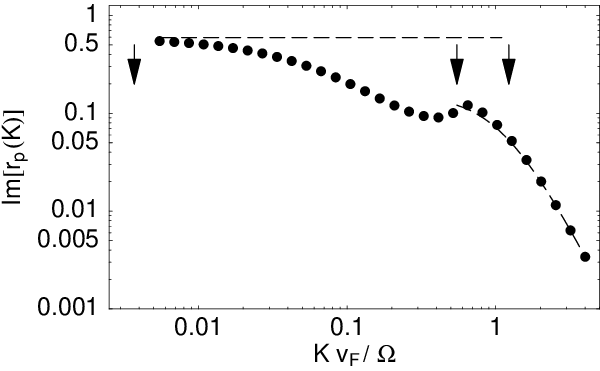}%
\hspace*{-40mm}%
\caption{Reflection coefficient ${\rm Im}\,r_{\rm p}( K )$ vs. the
normalized wave vector $K v_F / \Omega$. Dashed lines:
asymptotical formulas~(\ref{eq:r-p-asymp}). 
Symbols: numerical calculation.
The arrows mark, from left to right, the characteristic scales 
$\omega / c$, $1/|\ell|$ and $1/\Lambda$. Chosen parameters: 
$\Omega / \nu = 192$, $c / v_{F} = 148$, ($v_{F}/\Omega = 
0.84\,{\rm \AA}$), taken from the Drude description of aluminium
\cite{AshcroftMermin}. We take $\varepsilon_{b} = 2$
to model the contribution of bound electrons. 
Left panel: $\omega = 0.8\,\nu$ or $\lambda = 19\,\mu{\rm m}$.
Right panel: $\omega = 0.55\,\Omega$ ($\lambda = 140\,{\rm nm}$), 
near the large-$K$ asymptote of the 
surface plasmon resonance in the local approximation (given by 
$\varepsilon_{\rm loc} + 1 = 0$).
}
\label{fig:rp-metal}
\end{figure}

The reflection coefficient is plotted in Fig.\ref{fig:rp-metal} where 
the asymptotic expressions~(\ref{eq:r-p-asymp}) are represented as 
dashed lines. We find good agreement outside the crossover range
$K |\ell| \sim 1$.
In the frequency range of the anomalous skin effect, $\omega \sim 
\nu$ (left panel, $\lambda = 19\,\mu{\rm m}$ in the infrared), 
the nonlocal plasma shows an increased 
${\rm Im}\,r_{\rm p}( K )$, with a cutoff occurring beyond
$K_{c} \sim 1/\Lambda$ [see Eq.(\ref{eq:dimless-g})]. This effect
is well known \cite{Ford84} and is related to the enhanced
spontaneous emission rate for a nonlocal metallic surface that
was recently pointed out \cite{Larkin04}.
The reflection loss remains small in absolute numbers because 
of the large conductivity of the material.
The opposite behaviour is found near the (local, non-retarded) 
surface plasmon resonance 
(right panel, $\lambda = 140\,{\rm nm}$ in the far UV): 
${\rm Im}\,r_{\rm p}( K )$ decreases from its local value, with a
weakly resonant feature emerging around $K \sim 1/|\ell|$.

From these plots, we observe that the characteristic wave vector 
scale $K_{c}$ strongly depends on the frequency range. An upper limit 
is set by $1/\Lambda$, involving the Thomas-Fermi screening length, 
but significant changes already occur on the scale $1/|\ell|$. 
The 
characteristic distance below which non-local effects become manifest, 
is thus given by the largest of $|\ell|$ and $\Lambda$. This is 
typically $|\ell|$, since in order 
of magnitude, $|\ell| / \Lambda \sim \Omega / |\omega + {\rm 
i}\nu|$ which is much larger than unity for good conductors up to 
the visible domain. At frequencies smaller (larger) than the damping rate 
$\nu$, 
the mean free path $l_{\rm mfp}$ 
(the ``ballistic amplitude'' $v_{F} / \omega$): 
sets the scale for nonlocal effects, respectively. 

We note that for typical metals, the Thomas-Fermi scale $\Lambda$ does 
not differ much from the Fermi wavelength $1/k_{F}$. The asymptotics
derived above within the semiclassical Lindhard model~(\ref{eq:eps-l-Lindhard}) 
is therefore only qualitatively valid at short distances (large
wavevectors).

\subsection{Energy density and lateral coherence}

The numerical calculation of the correlation function Eq.(\ref{eq:coh-Im-r}) 
can be done efficiently using a numerical interpolation of 
${\rm Im}\,r_{p}( K )$ that we continue  
for large and small $K$ using the asymptotics derived above.

\begin{figure}
    \includegraphics*[width=70mm]{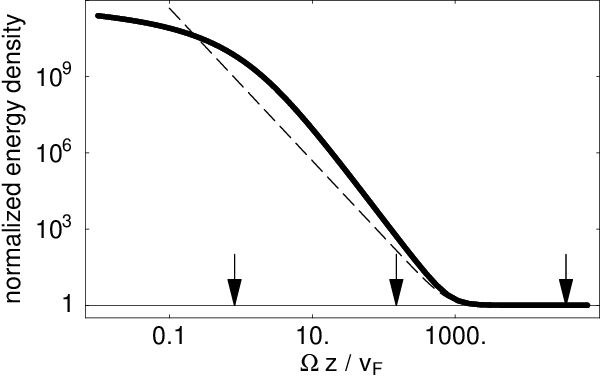}
    \hspace*{5mm}
    \includegraphics*[width=70mm]{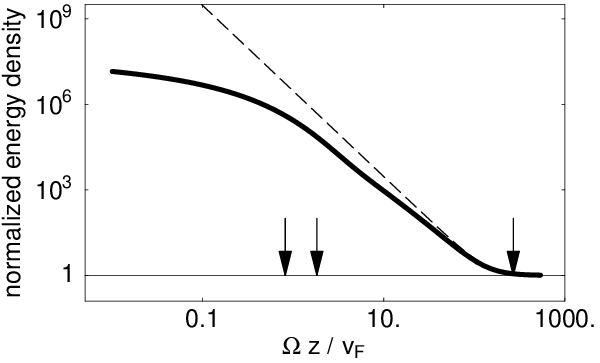}%
\hspace*{-40mm}%
\caption{Electric energy density, normalized to its far-field
value, vs.\ normalized distance $\Omega z / v_F$.
Dashed line: local dielectric.
Solid line: numerical calculation (left: $\omega = 0.8\,\nu$;
right: $\omega = 0.55\,\Omega$; other parameters as in 
Fig.\ref{fig:rp-metal}).
The arrows mark, from left to right, the characteristic distances
$\Lambda$, $|\ell|$, and $\lambda = 2\pi c / \omega$.}
\label{fig:ue-metal}
\end{figure}

We plot in Fig.\ref{fig:ue-metal} the electric energy density as a 
function of distance, for the same two frequencies as in 
Fig.\ref{fig:rp-metal}. 
Deviations from 
the local approximation (dashed line) occur at distances smaller than 
$|\ell|$: enhancement at low frequencies ($\omega \sim \nu$, left 
panel), suppression near the surface plasmon 
resonance (right panel), which is consistent
with the trends seen in Fig.\ref{fig:rp-metal}. A saturation at small 
distances is also visible, although it occurs for fairly small values 
of $\Omega z / v_{F}$ (where the semi-classical Lindhard function is 
in practice no longer valid). We note also that for $z \ge \lambda$, the 
plots are only qualitative since the calculation does not take into 
account retardation.
%
% and cutoff, mark $z = 1/K_{c}$. Two extreme 
% parameters: $|\omega + {\rm i}\nu| \gg$ and $\ll \Omega$.

\begin{figure}
\includegraphics*[width=80mm]{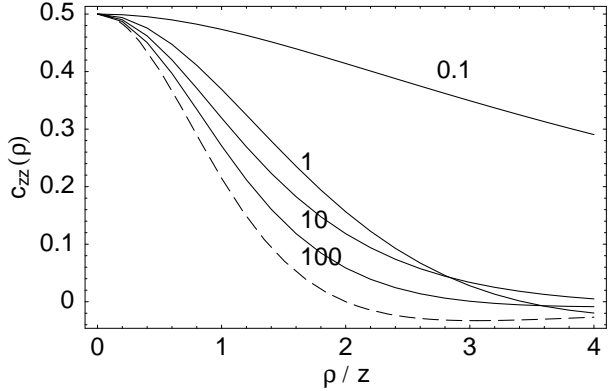}
\caption{Normalized degree of spatial coherence for $z$-polarized 
fields, probed at a lateral separation $\rho$.
The numbers on the curves (solid lines) give the normalized distance 
$\Omega z / v_F = 100, 10, 1, 0.1$, with the normalized
Thomas-Fermi screening length being $\Omega \Lambda / v_F = 
(\varepsilon_{b} / 3)^{1/2} \approx 0.8$. Dashed line: result for a
local dielectric in the near-field limit $z \ll \lambda$, 
taken from Eq.(\ref{eq:local-xnf}).
The chosen parameters are those of Fig.\ref{fig:rp-metal}, right panel.}
\label{fig:czz}
\end{figure}

Finally, we illustrate the finiteness of the coherence length as the 
distance of observation enters the nonlocal regime. We plot in 
Fig.\ref{fig:czz} the $zz$-component of the normalized coherence 
tensor~(\ref{eq:degree-of-coherence}), as a function of the lateral 
separation $\rho / z$. In the local regime, one gets a universal 
curve independent of the distance (dashed line). 
This is no longer true near a 
nonlocal metal: when Thomas-Fermi screening sets in ($z \le \Lambda$), 
the coherence function departs from its local limit, its width 
(the coherence length) becoming much larger than $z$. 
% 
% Plot coherence function vs. $s / z$ (numerically computed): deviation 
% from this scaling when $z \ll 1/K_{c}$. 

\section{Concluding remarks}

We have discussed in this paper the impact of a nonlocal dielectric
response on the spatial coherence of thermal electromagnetic near 
fields above a planar surface. Using two different models to
describe the nonlocal response, we have shown that when the 
sources of the field have a finite correlation length, this length
sets the minimum scale for the coherence length of the field as well.
This behaviour is qualitatively similar to what we found previously
when investigating the contribution of thermally excited surface
plasmons where coherence length and plasmon propagation length
coincide \cite{Henkel00b}. 
We have thus provided semi-quantitative evidence for the impact
of nonlocality that has been conjectured already by Rytov's group
\cite{Rytov3}.

The calculation for an electron plasma model highlights, on the
one hand, the crucial role played by Thomas-Fermi screening,
that sets the minimum coherence length.
On the other hand, significant deviations from the local
description already occur at scales below the electron mean free 
path (Fig.\ref{fig:rp-metal} and Fig.\ref{fig:ue-metal}), although
these are not accompanied by an increase in spatial coherence. 

Our calculations can be improved taking into account quantum effects
in the Lindhard dielectric function \cite{Ford84}, which will lead
to quantitative changes at short distance. Indeed, for typical metals,
the Thomas-Fermi screening length $v_F/\Omega$ and the Fermi wavelength
$1/k_F$ are fairly close \cite{AshcroftMermin}. A comparison to other
models of nonlocal dielectric functions would be interesting as well. 
On the experimental side, it would be interesting to compare the recent 
data on heat transfer between a scanning tip and a surface
\cite{Kittel05} with a microscopic calculation along the lines used
here. We also mention that in the context of the
Casimir force, nonlocal surface impedances have been studied.
The nonlocal correction is particularly relevant at finite temperature
and large distances and leads to a behaviour of the Casimir force
that is qualitatively similar, even without absorption, 
to the local, lossy Drude model, see for example 
Refs.\cite{Svetovoy05a,Sernelius05a}. Finally, it remains
to study the impact of another property of real metals, the smooth
rather than abrupt transition of the electron density profile at the 
surface: this can be described by effective surface displacements
that depend on both polarization and wave vector,
thus adding to the nonlocal effects considered here
\cite{Feibelman82}.

\smallskip\

{\footnotesize\noindent
We thank R\'emi Carminati and Jean-Jacques Greffet for discussion
and Illarion Dorofeyev and Francesco Intravaia for helpful comments.
C.H.\ acknowledges support from the European Commission (network FASTNet
and projects ACQP and QUELE).

}

\appendix

\section{Appendix}

Les us consider the Green tensor relating an electric current in 
a local medium 2 ($z'<0$) to the electric field in medium 1 ($z>0$) that we 
take as vacuum ($\varepsilon_{1} = 1$). 
This tensor can be written as an expansion in plane waves (Weyl
expansion)
\begin{equation}
\label{ }
G_{ij}({\bf r},{\bf r'}) = 
\frac{ {\rm i} }{ 2 }
\int \frac{ \mathrm{d}^2{\bf K} }{ (2\pi)^2 \,\gamma_{2} } 
g_{ij}({\bf K})e^{i [ k_{x} (x - x') + k_{y} (y - y') ]}
e^{i \gamma_1 z} e^{-i \gamma_2 z'},
\end{equation}
where ${\bf K} = (k_{x}, k_{y})$ is the wave vector component parallel to the
interface.  The $\gamma_i$ are the $z$-components of the wave vector:
$\gamma_i^2=\epsilon_i(\omega/c)^2 - K^2$.  In the notation of 
Ref.\cite{Henkel00b},
\begin{equation}
\label{ }
g_{ij}({\bf K})=\sum_{\mu=s,p}e_{\mu,i}^{(t)}e_{\mu,j}^{(2)}t_\mu^{21}
\end{equation}
The polarization vectors for the $s$ and $p$ polarization are
\begin{eqnarray}
{\bf e}^{(t)}_{\rm s} & = & 
{\bf e}^{(2)}_{\rm s} = 
\hat{\bf K} \times \hat{\bf e}_z
\\
{\bf e}^{(t)}_{\rm p} & = & 
\frac{ K \hat{\bf z} - \gamma \hat{\bf K} }{  
\omega / c  }
\\
{\bf e}^{(2)}_{\rm p} & = & 
\frac{ K \hat{\bf z} - \gamma_2 \hat{\bf K} }{ \sqrt{\varepsilon_{2}} 
\,\omega / c  }
\label{eqa:pol-vectors}
\end{eqnarray}
where $\hat{\bf K}$ is the unit vector parallel to ${\bf K}$.
The $t_\mu^{21}$ are the Fresnel transmission coefficients between media 2
and 1:
\begin{eqnarray} 
t_{\rm s}^{21} & = &
\frac{ 2 \gamma_2 }{
\gamma_{1} + \gamma_2 },
\qquad
t_{\rm p}^{21} =
\frac{ 2 \gamma_2 \sqrt{ \varepsilon_{2} } }{
\varepsilon_{2} \gamma_{1} + \gamma_2 }
.
\label{eqa:def-t-sp}
\end{eqnarray}

  \newcommand{\mybstpath}{/Users/carstenh/Biblio/Database/bst/}
  \newcommand{\mybibpath}{/Users/carstenh/Biblio/Database/}

  \bibliography{\mybibpath journals,%
  \mybibpath bib-ac,\mybibpath bib-dh,%
  \mybibpath bib-io,\mybibpath bib-pz,%
  \mybibpath bib-2004,\mybibpath bib-2005}

\begin{thebibliography}{10}

\bibitem{Planck00}
M. Planck: Verh. Dt. phys. Ges. (Berlin) {\bf 2}, 237 (1900)

\bibitem{Rytov3}
S.~M. Rytov, Y.~A. Kravtsov, V.~I. Tatarskii: {\em Elements of Random Fields},
  Vol.~3 of {\em Principles of Statistical Radiophysics}. Berlin: Springer 1989

\bibitem{Gori94}
F. Gori, D. Ambrosini, V. Bagini: Opt. Commun. {\bf 107}, 331 (1994)

\bibitem{Greffet99}
R. Carminati, J.-J. Greffet: Phys. Rev. Lett. {\bf 82}, 1660 (1999)

\bibitem{Henkel00b}
C. Henkel, K. Joulain, R. Carminati, J.-J. Greffet: Opt. Commun. {\bf 186}, 57
  (2000)

\bibitem{Savasta99}
O.~D. Stefano, S. Savasta, R. Girlanda: Phys. Rev. A {\bf 60}, 1614 (1999)

\bibitem{Chance78}
R.~R. Chance, A. Prock, R. Silbey:  in {\em Advances in Chemical Physics
  XXXVII}, edited by I. Prigogine, S.~A. Rice. New York: Wiley \& Sons 1978,
  pp.\ 1--65

\bibitem{Dunn99}
R.~C. Dunn: Chem. Rev. {\bf 99}, 2891 (1999)

\bibitem{Mohideen02}
F. Chen, U. Mohideen, G.~L. Klimchitskaya, V.~M. Mostepanenko: Phys. Rev. Lett.
  {\bf 88}, 101801 (2002)

\bibitem{Xu94}
J.-B. Xu, K. Lauger, R. Moller, K. Dransfeld, I.~H. Wilson: J. Appl. Phys. {\bf
  76}, 7209 (1994)

\bibitem{Pendry99b}
J.~B. Pendry: J. Phys. Cond. Matt. {\bf 11}, 6621 (1999)

\bibitem{Mulet01}
J.-P. Mulet, K. Joulain, R. Carminati, J.-J. Greffet: Appl. Phys. Lett. {\bf
  78}, 2931 (2001)

\bibitem{Kittel05}
A. Kittel, W. M\"uller-Hirsch, J. Parisi, S.-A. Biehs, D. Reddig, M. Holthaus:
  Phys. Rev. Lett. {\bf 95}, 224301 (2005)

\bibitem{MandelWolf}
L. Mandel, E. Wolf: {\em Optical coherence and quantum optics}. Cambridge:
  Cambridge University Press 1995

\bibitem{Joulain03}
K. Joulain, R. Carminati, J.-P. Mulet, J.-J. Greffet: Phys. Rev. B {\bf 68},
  245405 (2003)

\bibitem{Friberg02}
T. Set{\"a}l{\"a}, M. Kaivola, A.~T. Friberg: Phys. Rev. Lett. {\bf 88}, 123902
  (2002)

\bibitem{Wolf04a}
J. Ellis, A. Dogariu, S. Ponomarenko, E. Wolf: Opt. Lett. {\bf 29}, 1536 (2004)

\bibitem{Girard00a}
C. Girard, C. Joachim, S. Gauthier: Rep. Prog. Phys. {\bf 63}, 893 (2000)

\bibitem{HenkelHabil}
C. Henkel: {\em Coherence theory of atomic de Broglie waves and electromagnetic
  near fields}. Potsdam: Universit{\"a}tsverlag 2004, online at
  http://opus.kobv.de/ubp/volltexte/2005/135/

\bibitem{Scheel99a}
S. Scheel, L. Kn{\"o}ll, D.-G. Welsch: acta phys. slov. {\bf 49}, 585 (1999)
  [quant-ph/9905007].

\bibitem{Polder71}
D. Polder, M.~V. Hove: Phys. Rev. B {\bf 4}, 3303 (1971)

\bibitem{Henry96}
C.~H. Henry, R.~F. Kazarinov: Rev. Mod. Phys. {\bf 68}, 801 (1996)

\bibitem{Callen51}
H.~B. Callen, T.~A. Welton: Phys. Rev. {\bf 83}, 34 (1951)

\bibitem{Eckhardt82}
W. Eckhardt: Opt. Commun. {\bf 41}, 305 (1982)

\bibitem{KliewerFuchs}
K.~L. Kliewer, R. Fuchs: Adv. Chem. Phys. {\bf 27}, 355  (1974)

\bibitem{Sipe84}
J.~M. Wylie, J.~E. Sipe: Phys. Rev. A {\bf 30}, 1185 (1984)

\bibitem{Abramowitz}
{\em Handbook of Mathematical Functions}, ninth ed., edited by M. Abramowitz,
  I.~A. Stegun. New York: Dover Publications, Inc. 1972

\bibitem{Palik}
{\em Handbook of optical constants of solids}, edited by E. Palik. San Diego:
  Academic 1985

\bibitem{AshcroftMermin}
N.~W. Ashcroft, N.~D. Mermin: {\em Solid State Physics}. Philadelphia: Saunders
  1976

\bibitem{Ford84}
G.~W. Ford, W.~H. Weber: Phys. Rep. {\bf 113}, 195 (1984)

\bibitem{Kliewer68}
K.~L. Kliewer, R. Fuchs: Phys. Rev. {\bf 172}, 607 (1968)

\bibitem{Agarwal75a}
G.~S. Agarwal: Phys. Rev. A {\bf 11}, 230 (1975)

\bibitem{Dorofeyev02a}
I. Dorofeyev, H. Fuchs, J. Jersch: Phys. Rev. E {\bf 65}, 026610 (2002)

\bibitem{Larkin04}
I.~A. Larkin, M.~I. Stockman, M. Achermann, V.~I. Klimov: Phys. Rev. B {\bf
  69}, 121403(R) (2004)

\bibitem{Svetovoy05a}
V.~B. Svetovoy, R. Esquivel: Phys. Rev. E {\bf 72}, 036113 (2005)

\bibitem{Sernelius05a}
B.~E. Sernelius: Phys. Rev. B {\bf 71}, 235114 (2005)

\bibitem{Feibelman82}
P.~J. Feibelman: Progr. Surf. Sci. {\bf 12}, 287 (1982)

\end{thebibliography}
  \bibliographystyle{\mybstpath springer}

\end{document}